\documentclass[letterpaper,10pt,leqno]{article}
\usepackage{xchk}

\hypersetup{
  pdftitle={xChk: Bring Your Own Identity --- Heterogeneous Assurance with Verifier-Determined Sufficiency},
  pdfauthor={Sean MacGuire},
  pdfkeywords={Bring Your Own Identity, BYOI, Identity Assurance, OAuth/OIDC, OpenID Connect, Agent Attestation, Identity Provider, Delegation, Portfolio Claims, Human-in-the-Loop}
}

\begin{document}

\title{xChk: Bring Your Own Identity --- Heterogeneous Assurance with Verifier-Determined Sufficiency}
\author{Sean MacGuire\\[0.35em]{\mdseries\normalsize Independent Researcher}\\[0.2em]{\small \texttt{sean@maclawran.ca}}}
\date{July 2026}

\maketitle

\begin{abstract}
We present xChk, a reference identity provider for Bring Your Own Identity (BYOI): users enroll via heterogeneous proofs (government KYC, corporate SSO, WebAuthn/FIDO2, professional networks, live verification, longitudinal activity, behavioral signals) and disclose them as portfolio claims in standard OAuth 2.0 / OpenID Connect (OIDC) tokens, while each relying party applies its own sufficiency policy---the IdP transports claims and may evaluate an RP-supplied evidence policy for consent, but does not adjudicate access. Enrollment depth varies by modality (some paths are user-initiated; org KYB and officer binding are operator-assisted).

\par
xChk also supports human-in-the-loop attestation for high-risk actions: humans can initiate attestations directly (browser UI / \texttt{POST /api/attestations}), and AI agents acting under those principals can trigger the same gateway via scope-gated authorize/attest---hash-chained human approvals on a shared verification graph (humans via OIDC; agents via API keys). A production deployment at \url{https://in.xchk.io} ships both initiation paths with bilateral RP evaluation at consent; one documented relying party (\url{https://crabbyed.com}, Appendix~\ref{s:crabbyed}) exercises Login with xChk.
\end{abstract}

\keywords{Bring Your Own Identity $\cdot$ BYOI $\cdot$ Identity Assurance $\cdot$ OAuth/OIDC $\cdot$ OpenID Connect $\cdot$ Agent Attestation $\cdot$ Identity Provider $\cdot$ Delegation $\cdot$ Portfolio Claims $\cdot$ Human-in-the-Loop}

\section{Introduction}\label{s:introduction}

Identity providers (IdPs) such as Okta, Auth0, Microsoft Entra ID, and Google Identity Platform authenticate users and issue tokens enabling access to third-party applications via OAuth 2.0~\citep{rfc6749} and OpenID Connect (OIDC)~\citep{oidccore}. These systems share a fundamental limitation: they authenticate via a single modality (password, social login, or SAML federation) and issue a token asserting only that the user authenticated---not what level of identity assurance was established. A user authenticated via corporate SSO is indistinguishable from a user who has provided a government-issued ID and completed live-face verification, despite the latter representing a far higher degree of assurance.

Specialized identity verification platforms (Persona, Jumio, Onfido, Didit) perform KYC checks including government ID verification and liveness detection, but do not function as general-purpose OAuth/OIDC identity providers. Decentralized identity systems (W3C Verifiable Credentials~\citep{w3cvc}, Microsoft Entra Verified ID, cheqd, Polygon ID) issue verifiable credentials but require relying parties to implement VC verification infrastructure and do not integrate with the OAuth/OIDC flow used by thousands of existing applications. Machine identity systems (SPIFFE/SPIRE~\citep{spiffe}, cloud IAM roles) operate in a separate domain from human identity federation and cannot convey the human operator behind a deployed agent.

Meanwhile, AI agents are increasingly performing consequential actions---financial transactions, file system mutations, external API calls---on behalf of human principals~\citep{kampik2022governance, schultze2025building}. These agents operate with increasing autonomy, and existing governance frameworks~\citep{wang2026agentspec} focus on behavioral compliance but assume the identity and authorization layers are solved. AgentBound~\citep{agentbound}, a complementary runtime governance framework, evaluates whether an authorized action should execute under current behavioral context, but does not address how the agent's identity or its human principal's identity assurance is established.

\textbf{The gap.} To our knowledge, no production OAuth/OIDC identity provider combines BYOI multi-modality enrollment, portable portfolio claims with verifier-determined sufficiency, unified human/non-human governance on one verification graph, human-rooted delegation with cascading revocation, and human-in-the-loop attestation with optional blockchain anchoring. We detail this gap and xChk's shipped coverage in \S\ref{s:gap}.

\paragraph{Contributions.}
\begin{enumerate}
  \item \textbf{Portfolio claims in OIDC with bilateral sufficiency}---heterogeneous enrollment modalities encoded as typed portfolio artifacts in standard OAuth/OIDC tokens; the IdP transports claims and may evaluate an RP-supplied evidence policy for consent, while the RP adjudicates access; production adds bilateral evaluation of the RP's institutional portfolio before subject portfolio release (\S\ref{s:byoi}).
  \item \textbf{Human-rooted attestation on one verification graph}---humans and agents share org enrollment, portfolio assembly, and attestation records; humans federate via Login with xChk (surface~A) and may initiate attestations directly; agents use API keys with authorize/attest, scope-monotonic delegation, cascading revocation, and hash-chained human approvals (surfaces B/C; \S\ref{s:token}, \S\ref{s:attestation}, \S\ref{s:surfaces}).
  \item \textbf{Production reference implementation} at \url{https://in.xchk.io} with one documented RP (\url{https://crabbyed.com}, Appendix~\ref{s:crabbyed}), ten portfolio artifact types, Hermes guardrails E2E integration, a public portfolio HMAC verifier (Appendix~\ref{s:hmac-verifier}), and a micro-benchmark of public endpoints (\S\ref{s:microbench}).
\end{enumerate}

\section{Related Work}\label{s:related}

\subsection{Identity Federation and OAuth/OIDC Providers}

Traditional IdPs (Okta, Auth0, Microsoft Entra ID, Google Identity Platform) authenticate users through a single modality (password, social login, or SAML federation) and assert only that authentication occurred. None encode heterogeneous identity assurance levels as standard OIDC claims. Step-up authentication and risk-based authentication~\citep{osho2021risk} modulate the authentication challenge based on risk signals but do not produce a token carrying the user's full identity assurance profile for independent RP evaluation.

\subsection{KYC and Identity Verification Platforms}

Persona, Jumio, Onfido, and Didit perform government ID checks and liveness detection but issue verification results (boolean flags, reports, or W3C VCs)---not OAuth/OIDC tokens. A relying party using these platforms must implement separate integration, verification, and storage for each platform's output. xChk integrates these as enrollment modalities within a standard OAuth/OIDC flow.

\subsection{Decentralized Identity (W3C VCs, DIDs)}

Verifiable Credentials~\citep{w3cvc} and Decentralized Identifiers (DIDs)~\citep{w3cdid} enable cryptographically verifiable claims but require relying parties to implement VC verification infrastructure (DID resolution, credential validation, schema parsing) not present in standard OAuth/OIDC. Microsoft Entra Verified ID, cheqd, and Polygon ID issue VCs for specific claims but typically support one credential type per issuer rather than aggregating multiple heterogeneous identity proofs into a single token.

\subsection{Machine Identity}

SPIFFE/SPIRE~\citep{spiffe} issues X.509-SVIDs for workload identity. Cloud IAM roles (AWS IAM, GCP IAM, Azure Managed Identities) provide identity documents for services. HashiCorp Vault issues short-lived tokens. These systems cannot convey the human operator who deployed a workload or the delegation chain authorizing an agent's actions, and they operate outside the OAuth/OIDC protocol family.

\subsection{Agent Identity and Authorization}

The OpenID Foundation has proposed OpenID Connect for Agents (OIDC-A) 1.0~\citep{oidca}, an extension to OIDC for representing, authenticating, and authorizing LLM-based agents. OIDC-A defines agent identity claims, delegation chain semantics, and attestation verification within the OAuth 2.0 ecosystem. xChk differs from OIDC-A in that xChk's identity provider is BYOI-native---it accepts enrollment via multiple independent human identity modalities and produces tokens carrying human assurance portfolios, while agent governance is practiced through a parallel authorize/attest API that can converge toward OIDC-A-compatible delegation claims. The two approaches are complementary: a relying party could accept xChk human portfolio tokens and OIDC-A agent claims in a single policy engine.

Forter's WO2026039510A1~\citep{forter} describes agent authentication protocols for payment/commerce domains using proprietary certificates---domain-specific, not standard OAuth/OIDC. Onesource's US20250373432A1~\citep{onesource} describes compliance-token inheritance for healthcare with custom tokens---domain-specific, not general-purpose identity provider functionality.

Microsoft's Agent Governance Toolkit (AGT)~\citep{microsoftagt} provides in-process policy middleware with \texttt{require\_approval} hooks and local audit trails. AGT does not issue OAuth/OIDC portfolio tokens, perform multi-modality human enrollment, or hash-chained passkey attestation off the agent runtime; xChk composes with AGT via an adapter that routes approval obligations to the attestation gateway.

Context Lineage Assurance for Non-Human Identities~\citep{singh2025context} addresses cryptographic lineage for A2A interactions but focuses on cryptographic proof chains rather than multi-source identity enrollment. AgentRiskBOM~\citep{dutta2025agentriskbom} defines a security BOM for scoping agent risk (autonomy, tool permissions, approval gates) but is a metadata artifact, not an identity protocol. Agent registry surveys~\citep{singh2025registry} catalog MCP, A2A, Entra Agent ID, and NANDA/AgentFacts approaches---none provide BYOI with heterogeneous modality aggregation.

Separately, U.S.\ AI regulation is fragmenting along state lines~\citep{orrick2026ailaw}: California, Colorado, and dozens of other jurisdictions now impose distinct deployer and developer obligations (training-data transparency, automated decision-making, ``AI acted autonomously'' liability limits, content provenance, and sector rules). Existing agent registries and HITL products rarely treat \emph{registration jurisdiction} as a first-class attribute of a non-human identity that humans can read at approval time and that auditors can recover from the attestation record.

\subsection{Behavioral Governance}\label{s:agentbound}

AgentBound~\citep{agentbound} provides a runtime governance framework that evaluates each proposed agent action using three independent authorities: delegated authorization, owner-signed behavioral constitutions, and site action contracts. Their judgments are conservatively composed through a formal decision lattice (Deny $<$ Review $<$ Permit). AgentBound generates cryptographically verifiable governance receipts binding every action to the policy artifacts governing the decision. xChk is complementary to AgentBound: AgentBound assumes identity is established and governs \emph{behavioral} correctness; xChk establishes identity assurance and enables \emph{identity-layer} verification. A combined system would use xChk to establish the identity of the human principal and record human approvals, and AgentBound to govern whether individual actions comply with behavioral policy. AgentBound's review obligations (human approvals) can be satisfied through xChk's attestation gateway; completed approvals appear as portfolio artifacts and \texttt{xchk\_attestations} claims in refreshed tokens.

\subsection{Assurance Levels and Rich Authorization}

NIST SP 800-63~\citep{nist80063} and eIDAS~\citep{eidas} define discrete identity assurance / authentication levels (IAL/AAL, LoA) that relying parties consume as ordinal steps. xChk instead exposes a typed \textbf{portfolio} of modality-specific artifacts so an RP can require combinations (e.g.\ KYC + passkey + helpdesk) rather than a single LoA integer. OAuth 2.0 Rich Authorization Requests (RAR)~\citep{rfc9396} and GNAP~\citep{rfc9635} address fine-grained authorization payloads; they are complementary prior art for expressing what an RP wants authorized, not substitutes for multi-source identity enrollment into OIDC claims.

\subsection{Gap Analysis}\label{s:gap}

To our knowledge, no production system combines all of: (1) BYOI enrollment across independent modalities, (2) heterogeneous assurance levels as standard OAuth/OIDC portfolio claims, (3) verifier-determined sufficiency with the IdP assisting but not adjudicating RP access, (4) unified human + non-human entity governance on one verification graph, (5) delegation chains terminating at a verified human operator with cascading revocation, (6) attestation history carried forward via surface~B token embeds and \texttt{attestation\_completed} portfolio artifacts, and (7) blockchain-anchored human-in-the-loop attestation (human-initiated or agent-triggered). xChk ships (1)--(6) in production and (7) when Ravencoin anchoring is enabled; item (4) covers agent governance on surface C while RP OIDC tokens remain human-only today (\S\ref{s:surfaces}--\S\ref{s:impl-status}).

\section{Architectural Model}\label{s:model}

\begin{figure}[!ht]
  \centering
  \makebox[\linewidth][c]{\includegraphics[width=0.88\linewidth,height=0.48\textheight,keepaspectratio]{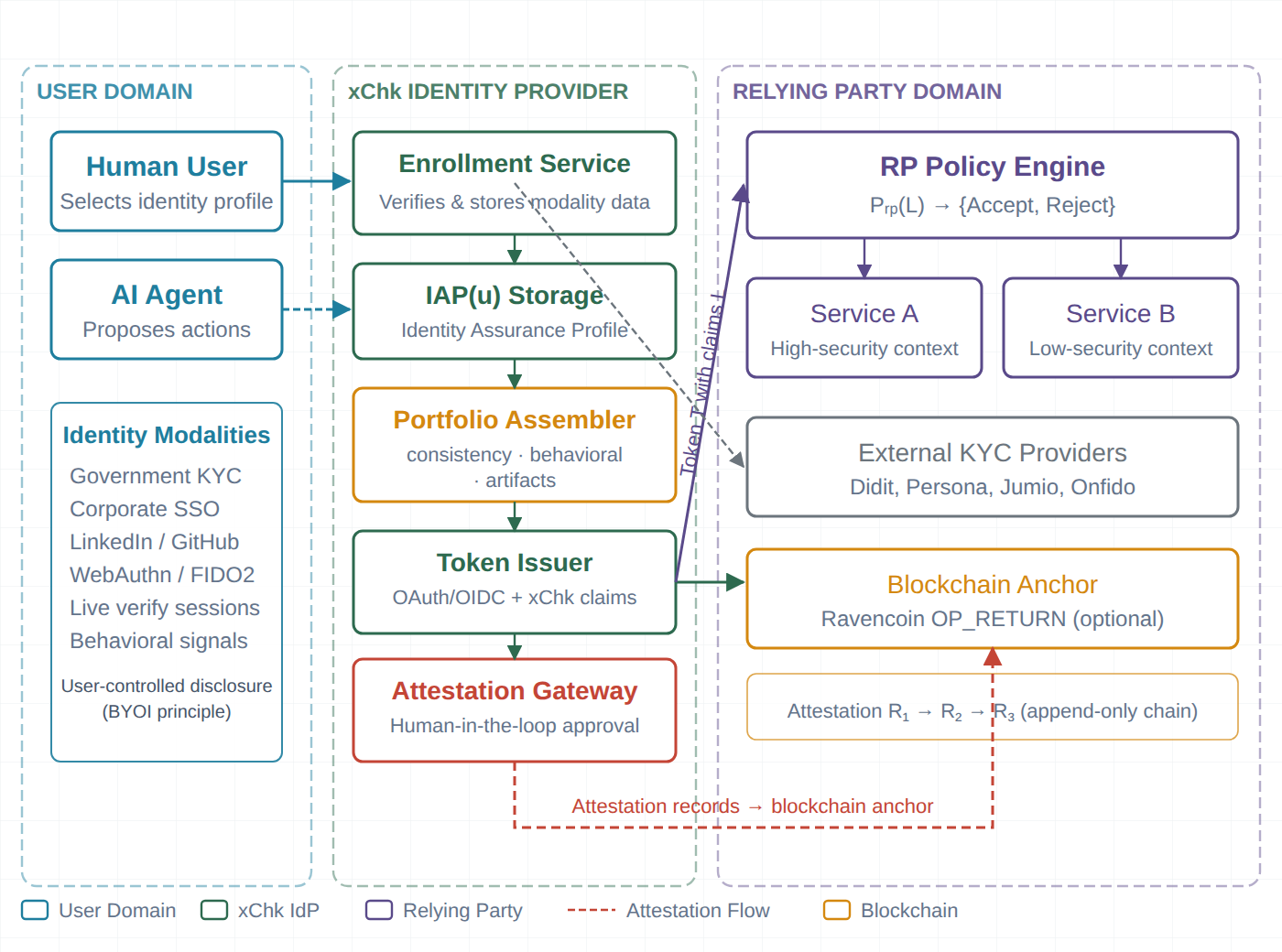}}
  \figcaption{xChk System Architecture}{Three-domain architecture showing the User Domain (modalities, agent), xChk Identity Provider (enrollment, IAP storage, portfolio assembly, token issuer, attestation gateway), and Relying Party Domain (policy engine, services, blockchain anchor).}
  \label{f:architecture}
\end{figure}

This section and \S\ref{s:attestation} give descriptive tuples and production mappings---an architectural model, not a security proof. The strongest structural properties we rely on are scope monotonicity down a delegation chain and cascading revocation of descendants (\S\ref{s:cascading}).

\subsection{Entities and Modalities}\label{s:entities}

Let $\mathcal{U}$ be the set of all users (human principals). Let $\mathcal{M} = \{m_1, \ldots, m_k\}$ be a finite set of identity verification modalities, where each modality $m \in \mathcal{M}$ is an enrollment channel that, upon successful completion, yields an assurance indicator $a(m)$ from a modality-specific assurance space $A_m$. Production implements these as typed portfolio artifacts (\S\ref{s:implementation}); examples include:

\begin{itemize}
  \item $A_{\mathit{kyc}} = \{\text{none}, \text{basic}, \text{govt\_id}\}$ --- government-issued identity verification (Didit)
  \item $A_{\mathit{sso}} = \{\text{none}, \text{tenant\_verified}\}$ --- corporate SSO (Entra ID, Google Workspace)
  \item $A_{\mathit{webAuthn}} = \{\text{none}, \text{single\_device}, \text{multi\_device}\}$ --- hardware-bound credentials (device biometric unlock at use time)
  \item $A_{\mathit{professional}} = \{\text{none}, \text{verified}\}$ --- professional network verification (LinkedIn)
  \item $A_{\mathit{live}} = \{\text{none}, \text{pass}, \text{fail}\}$ --- live helpdesk or interview verification
  \item $A_{\mathit{longitudinal}} = \{\text{none}, \text{present}\}$ --- platform activity over time (GitHub, LinkedIn)
  \item $A_{\mathit{behavioral}} = \{\text{usual}, \text{new}, \text{unusual}\}$ --- device, network, location, time-of-day consistency
  \item $A_{\mathit{attestation}} = \{\text{none}, \text{completed}\}$ --- prior human-in-the-loop approvals
\end{itemize}

For each user $u \in \mathcal{U}$, the identity provider maintains an \textbf{identity assurance profile}:

\begin{equation}
  \mathbf{IAP}(u) = \{ (m, a(m), t(m), p(m)) \;|\; m \in \mathcal{M}_{\mathit{enrolled}}(u) \}\label{e:iap}
\end{equation}

where $t(m)$ is the timestamp of modality $m$'s most recent verification and $p(m)$ is an optional provider identifier (including \texttt{issuer} and \texttt{issuerTier} for institutional provenance: \texttt{unverified}, \texttt{domain\_verified}, \texttt{entity\_verified}, \texttt{officer\_bound}). The set $\mathcal{M}_{\mathit{enrolled}}(u) \subseteq \mathcal{M}$ is the subset of modalities the user has chosen to complete---the BYOI principle.

\subsection{Identity Token Definition}\label{s:token}

\begin{figure}[!ht]
  \centering
  \makebox[\linewidth][c]{\includegraphics[width=0.88\linewidth,height=0.48\textheight,keepaspectratio]{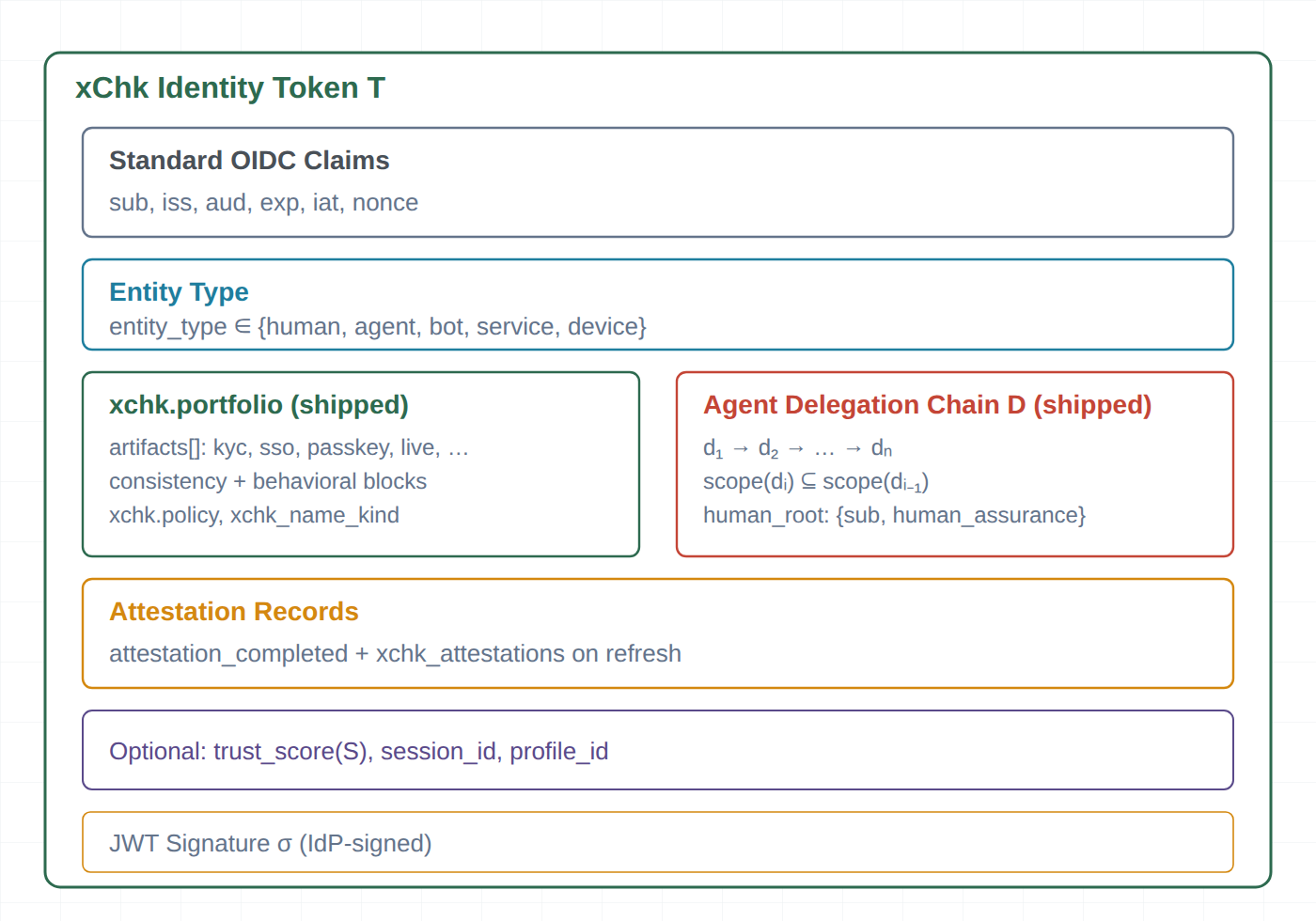}}
  \figcaption{xChk Identity Token Structure}{Standard OIDC claims, entity type, human assurance claims, agent delegation chain, attestation history, and advisory score. Production tokens include nested \texttt{xchk.portfolio} claims; \texttt{xchk\_entity\_type} is top-level; the access-token path adds \texttt{xchk\_trust\_score}, \texttt{xchk\_portfolio\_summary}, \texttt{xchk\_delegation\_chains}, and \texttt{xchk\_attestations}.}
  \label{f:token}
\end{figure}

An xChk identity token $\mathbf{T}$ for user $u$ at time $\tau$ is a tuple:

\begin{equation}
  \mathbf{T} = (\mathit{sub}, \mathit{iss}, \mathit{aud}, \mathit{exp}, \mathit{iat}, \mathit{entity\_type}, \mathbf{L}, \sigma)\label{e:token}
\end{equation}

where:
\begin{itemize}
  \item $\mathit{entity\_type} \in \{\text{human}, \text{agent}, \text{bot}, \text{service}, \text{device}\}$ --- entity-type taxonomy; production OIDC ID tokens issue \texttt{xchk\_entity\_type: human}; non-human entities are governed via the agent API (\S\ref{s:surfaces})
  \item $\mathbf{L} = \{ (m, a(m)) \;|\; m \in \mathcal{M}_{\mathit{enrolled}}(u) \land a(m) \neq \text{none} \}$ is the set of identity assurance claims
  \item $\sigma$ is the IdP's JWT signature over the token body
\end{itemize}

Production ID tokens (surface A) additionally carry a portfolio HMAC $\sigma_p$ over the consent-time package $\{\mathit{portfolio}, \mathit{policy}, \mathit{rpInstitutional}, \mathit{subjectRpPolicy}\}$ (wire encoding below); RS256 validation of $\sigma$ alone does not establish portfolio integrity.

For \textbf{agent} entity types, $\mathbf{L}$ additionally contains a \textbf{delegation chain} $\mathbf{D} = [d_1, \ldots, d_n]$ where each delegation link $d_i = (\mathit{entity\_id}, \mathit{scope}, \mathit{parent\_id}, \mathit{attestation\_id})$ satisfies:
\begin{itemize}
  \item $d_1.\mathit{parent\_id}$ is a human user $u \in \mathcal{U}$ (the root principal)
  \item For $i > 1$, $d_i.\mathit{parent\_id} = d_{i-1}.\mathit{entity\_id}$
  \item $\mathit{scope}(d_i) \subseteq \mathit{scope}(d_{i-1})$ (scope is monotonically non-expanding down the chain)
  \item Each $d_i$ references an attestation record (\S\ref{s:attestation-record}) proving the parent authorized the delegation
\end{itemize}

In production, surface~A OIDC ID tokens for human principals do not embed $\mathbf{D}$. When integrators supply \texttt{delegationChain} metadata at attestation create, completed attestations may surface $\mathbf{D}$ on surface~B via \texttt{xchk\_delegation\_chains} assembled from portfolio artifacts (\S\ref{s:cascading}); the platform does not auto-walk the \texttt{parentAgentId} agent tree to mint chains.

The token also optionally includes:
\begin{itemize}
  \item \path{xchk_trust_score}($S$): advisory composite $S$ on xChk access tokens
  \item \path{xchk_attestations}: ordered attestation records via \texttt{POST}~\path{/oauth/refresh-token} and as \path{attestation_completed} portfolio artifacts
  \item \path{xchk_session}: session identifier for multi-entity workflows via \texttt{POST}~\path{/api/session-tokens} and refresh-token embed
  \item \path{xchk_portfolio_summary}: time-bucketed artifact counts split by human and agent origin on access tokens and refresh
  \item Per-app artifact picker: user selects which artifact types each RP receives
\end{itemize}

\paragraph{Production OIDC encoding (v1).} Production ID tokens (scope \texttt{xchk\_portfolio}) expose:
\begin{itemize}
  \item \texttt{xchk.portfolio}---\texttt{artifacts[]} with typed records, plus \texttt{consistency} (volume, longevity, pass rates) and \texttt{behavioral} (usual/new/unusual per dimension)
  \item \texttt{xchk.policy}---result of evaluating the RP's \texttt{xchk\_evidence\_policy} (or client-registered policy): \texttt{\{ satisfied, missing, \ldots \}}
  \item \texttt{xchk.signature} / \texttt{xchk\_signature}---HMAC-SHA256 over the canonical consent-time package $\{\mathit{portfolio}, \mathit{policy}, \mathit{rpInstitutional}, \mathit{subjectRpPolicy}\}$ (signing key id \texttt{xchk-portfolio-v1}; RS256 JWT validation alone is insufficient)
  \item \texttt{xchk.rp\_institutional} / \texttt{xchk.subject\_rp\_policy}---RP institutional portfolio snapshot and subject-policy evaluation at consent (bilateral evaluation, \S\ref{s:byoi})
  \item \texttt{xchk\_name\_kind}, \texttt{xchk\_legal\_name\_on\_file}---name provenance (scope \texttt{profile})
  \item \texttt{xchk\_entity\_type}---\texttt{human} on OIDC ID tokens when portfolio scope is present
  \item \texttt{sub}---pairwise per OAuth client (cross-RP correlation resistant)
\end{itemize}

\paragraph{Wire encoding (v1).} Dotted names above (\texttt{xchk.portfolio}, etc.) are \textbf{documentation notation} for fields inside a nested JWT claim \texttt{xchk}, not literal dotted JWT keys. Surface~A ID tokens and userinfo responses nest portfolio fields under \texttt{xchk} and duplicate \path{xchk_signature}, \path{xchk_signing_key_id}, and \path{xchk_entity_type} at the top level. Integrators read \path{idToken.xchk.portfolio} for the paper's \texttt{xchk.portfolio}. Surface~B (\path{/api/login}, \texttt{POST}~\path{/oauth/refresh-token}) uses flat top-level underscore claims (\path{xchk_trust_score}, \path{xchk_portfolio_summary}, \path{xchk_attestations}, \path{xchk_delegation_chains}) on the first-party \path{xchkAccessToken}. The HMAC canonical payload uses camelCase keys (\path{rpInstitutional}, \path{subjectRpPolicy}), distinct from JWT snake\_case field names.

{\footnotesize
\begin{center}
\begin{tabular}{@{}L{0.32\linewidth}L{0.58\linewidth}@{}}
  \toprule
  \textbf{Doc notation} & \textbf{JWT access} \\
  \midrule
  \texttt{xchk.portfolio} & \path{payload.xchk.portfolio} \\
  \texttt{xchk.policy} & \path{payload.xchk.policy} \\
  \texttt{xchk.signature} & \path{payload.xchk.signature} or top-level \path{payload.xchk_signature} \\
  \texttt{xchk.rp\_institutional} & \path{payload.xchk.rp_institutional} \\
  \texttt{xchk.subject\_rp\_policy} & \path{payload.xchk.subject_rp_policy} \\
  \texttt{xchk\_entity\_type} & top-level \path{payload.xchk_entity_type} only \\
  \bottomrule
\end{tabular}
\end{center}
}

Optional scope \path{xchk_portfolio_proofs} adds expiring, audience-bound proof URLs and \texttt{sha256} / \texttt{rvnTxId} on attestation artifacts for verification independent of the IdP read path.

\paragraph{Access-token path (surface B).} \path{/api/login} and \texttt{POST}~\path{/oauth/refresh-token} mint an \path{xchkAccessToken}---a first-party JWT distinct from the OIDC access token returned by \path{/oauth/token}---carrying \path{xchk_trust_score}, \path{xchk_portfolio_summary}, \path{xchk_delegation_chains}, \path{xchk_attestations}, and optional \path{xchk_session}. Integrators call refresh after attestation approval to pick up new records without a full OIDC re-login.

\subsection{BYOI Decision Problem}\label{s:byoi}

Let $\mathbf{T}$ be an xChk identity token presented to a relying party (RP). The RP maintains a policy function:

\begin{equation}
  P_{\mathit{rp}} : 2^{\mathcal{M} \times A} \rightarrow \{\text{Accept}, \text{Reject}\}\label{e:byoi}
\end{equation}

where the input is a subset of $(m, a(m))$ pairs from $\mathbf{T}$ (in production: the artifact types the user consented to disclose). Critically:
\begin{itemize}
  \item The IdP does \textbf{not adjudicate access}---it verifies enrollment, assembles claims, and may evaluate an RP-\emph{supplied} policy for display (\texttt{xchk.policy} on the consent screen)
  \item RP does not see the full $\mathbf{IAP}(u)$---only the subset $\mathbf{L} \subseteq \mathbf{IAP}(u)$ the user chooses to present (per-app artifact picker)
  \item The user may withhold artifact types; \texttt{xchk.policy.satisfied} reflects the disclosed subset
\end{itemize}

This defines a \textbf{three-way separation of concerns}:
\begin{enumerate}
  \item \textbf{User} controls which modalities to enroll and which artifacts to disclose per RP
  \item \textbf{IdP} verifies modality enrollment truthfully and transports claims (plus optional policy evaluation assist)
  \item \textbf{RP} evaluates sufficiency and grants or denies access
\end{enumerate}

This separation emphasizes \textbf{portable multi-modality assurance portfolios} over conventional federation, where tokens typically assert only that authentication occurred (often via a single modality) rather than a user-selected set of typed assurance artifacts for independent RP evaluation.

\paragraph{Bilateral evaluation (production).} When scope \texttt{xchk\_portfolio} is requested, evaluation is concurrent in both directions: the RP's evidence policy is evaluated against the subject's disclosed portfolio (\texttt{xchk.policy}), and a subject-defined policy is evaluated against the RP's \textbf{institutional verification portfolio} assembled from the OAuth client's owning organization (DNS domain control, KYB, officer binding, participation history). The platform default subject policy requires \texttt{domain\_control}; unsatisfied evaluation returns \texttt{403} (\path{rp_institutional_policy_failed}) and disables portfolio release on the consent screen. The IdP still does not grant or deny RP access---it gates what the subject chooses to share. Token claims \texttt{xchk.rp\_institutional} and \texttt{xchk.subject\_rp\_policy} are included in the portfolio HMAC payload.

\subsection{Trust Score}\label{s:trust}

An optional trust score $S$ may be computed as a weighted function of enrolled modalities:

\begin{equation}
  S = \phi(a(m_1), \ldots, a(m_k), t(m_1), \ldots, t(m_k), N_a)\label{e:trust}
\end{equation}

where $N_a$ is the number of attested actions and $\phi$ is a monotonic non-decreasing function. Production ships \texttt{xchk\_trust\_score} on surface~B access tokens only; RPs on surface~A more commonly evaluate raw portfolio artifacts and \texttt{xchk.policy}. No production RP documented here gates access on $S$ alone.

\section{Attestation Protocol}\label{s:attestation}

\begin{figure}[!ht]
  \centering
  \makebox[\linewidth][c]{\includegraphics[width=0.88\linewidth,height=0.48\textheight,keepaspectratio]{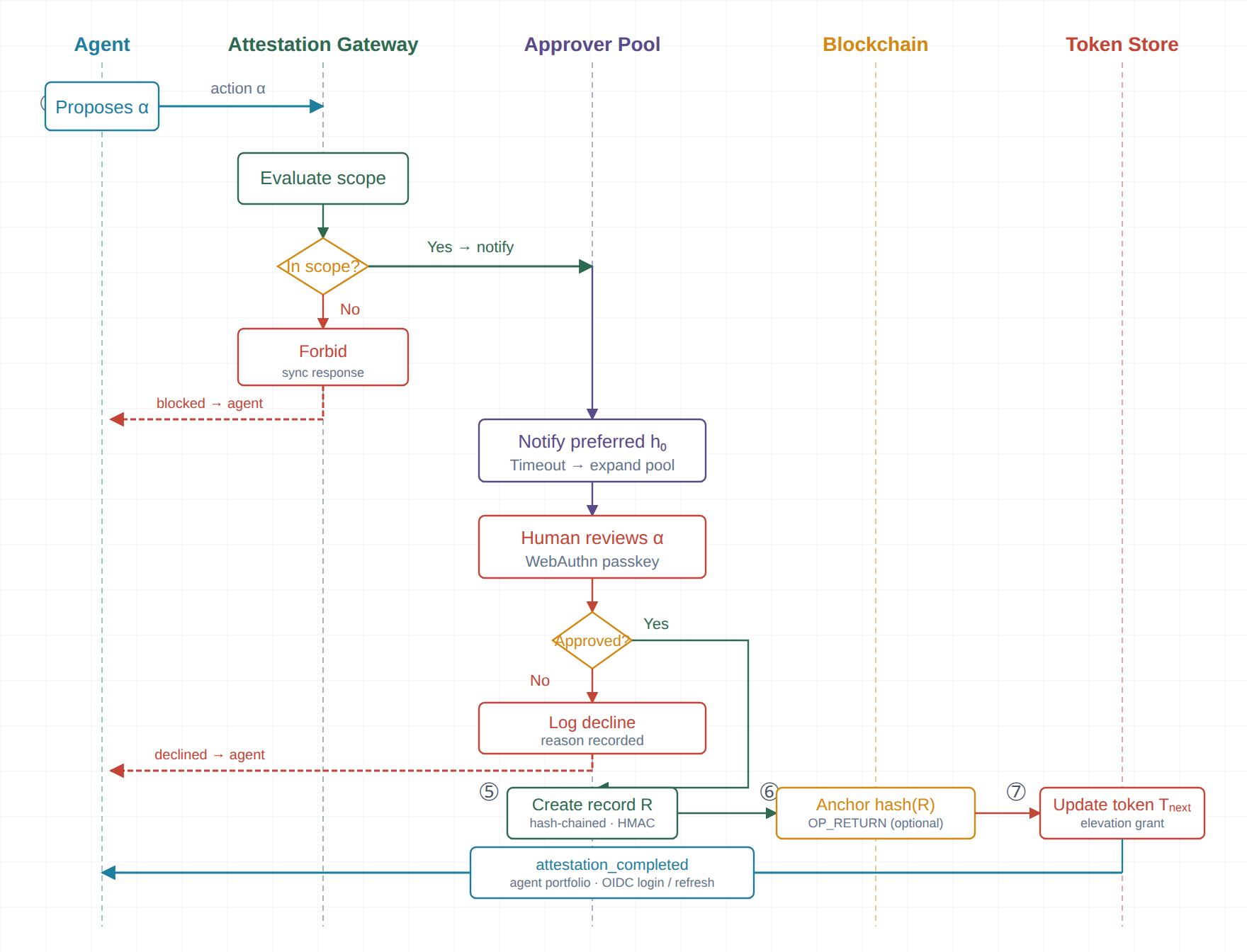}}
  \figcaption{Attestation Protocol Sequence}{Production attestation sequence (agent-triggered variant shown; humans may initiate via browser UI / \texttt{POST /api/attestations}): agent proposes action $\alpha$; gateway evaluates scope via \texttt{POST /api/agents/authorize}; progressive notification of approver pool; human approves/declines via WebAuthn passkey (required); attestation record $R$ created (hash-chained, HMAC-signed); optional Ravencoin OP\_RETURN anchor; short-lived elevation grant for retry; $R$ appears as \texttt{attestation\_completed} portfolio artifact on the approver's next OIDC login and via \texttt{xchk\_attestations} on token refresh.}
  \label{f:attestation}
\end{figure}

\subsection{Attestation Record}\label{s:attestation-record}

An attestation record $R$ is a tuple:

\begin{equation}
  R = (\mathit{att\_id}, \mathit{action}, \mathit{entity\_id}, \mathit{approver\_id}, \mathit{outcome}, \tau, \mathit{sig}, \mathit{chain\_hash})\label{e:record}
\end{equation}

where:
\begin{itemize}
  \item $\mathit{att\_id}$ --- globally unique identifier
  \item $\mathit{action} = (\mathit{type}, \mathit{target}, \mathit{args})$ --- the action being attested
  \item $\mathit{entity\_id}$ --- the entity that proposed the action (human principal or agent)
  \item $\mathit{approver\_id}$ --- the human approver's user identifier
  \item $\mathit{outcome} \in \{\text{approved}, \text{declined}, \text{escalated}\}$
  \item $\tau$ --- UTC timestamp
  \item $\mathit{sig}$ --- HMAC signature over the canonical record hash
  \item $\mathit{chain\_hash}$ --- SHA-256 hash of the previous attestation record in the chain (or null for the first)
\end{itemize}

Attestation records form an append-only chain. The hash of the latest attestation record may be written to a blockchain for decentralized timestamp anchoring when Ravencoin RPC is configured.

\subsection{Attestation Lifecycle (production)}\label{s:attestation-lifecycle}

The attestation gateway accepts human-initiated requests (\texttt{POST /api/attestations} / browser UI) and agent-triggered requests. When an agent entity $a$ proposes an action $\alpha$ that triggers a governance gate, the gateway executes:
\begin{enumerate}
  \item \textbf{Capture}: The agent runtime emits action $\alpha = (\mathit{type}, \mathit{target}, \mathit{args}, \mathit{context})$
  \item \textbf{Authorize}: \texttt{POST /api/agents/authorize} returns \texttt{allow}, \texttt{attest}, or \texttt{forbid}; \texttt{attest} mints a short-lived \texttt{authorizeSessionId} bound to the action hash
  \item \textbf{Approver resolution}: Policy rules select approver(s) $H \subseteq \mathcal{U}$ based on action type and severity; creation is rejected when \texttt{requiredApprovals} exceeds the available approver pool
  \item \textbf{Progressive notification}: Preferred approver $h_0$ is notified (Slack, Telegram, SMS, WhatsApp); on timeout $t$, the approver pool expands
  \item \textbf{Decision}: Approver $h \in H$ reviews $\alpha$, verifies via WebAuthn passkey, and produces $\mathit{outcome}$
  \item \textbf{Record creation}: Attestation record $R$ is created with hash chaining and HMAC signing; optional \texttt{delegationChain} metadata may be supplied by the integrator
  \item \textbf{Blockchain anchoring} (optional): $\mathit{hash}(R)$ written to Ravencoin OP\_RETURN when \texttt{RVN\_RPC\_PASS} is configured
  \item \textbf{Elevation grant}: On approval, an HMAC-signed grant ($\approx$15 min TTL, action-bound) allows the agent runtime to retry execution
  \item \textbf{Portfolio update}: $R$ surfaces as an \texttt{attestation\_completed} artifact on the human principal's next OIDC authorization and via \texttt{xchk\_attestations} on token refresh (\S\ref{s:token})
\end{enumerate}

Enterprise systems can consume completion webhooks---no production ERP/procurement adapter is shipped.

\subsection{Runtime Scope Enforcement (production)}\label{s:scope}

Agents authenticate with per-agent API keys, not OIDC tokens. Scope compliance is evaluated at authorize time: \texttt{POST /api/agents/authorize} classifies the action area, matches against the agent's registered grants, and returns \texttt{forbid} when out of scope. Out-of-scope actions may trigger attestation (escalation) rather than silent execution.

For delegation chain $\mathbf{D}$ embedded in token $\mathbf{T}$ or recorded on attestation create, $\mathit{scope\_compliant}(\alpha, \mathbf{D}) \iff \exists d_i \in \mathbf{D} : (\mathit{op}{:}r) \in \mathit{scope}(d_i)$. Expanded scope must satisfy $\mathit{scope}'(d_i) \subseteq \mathit{scope}(d_{i-1})$. Production evaluates scope at authorize time and embeds \texttt{xchk\_delegation\_chains} on xChk access tokens (surface B) from portfolio artifacts.

\subsection{Delegation Chains and Cascading Revocation}\label{s:cascading}

\begin{figure}[!ht]
  \centering
  \makebox[\linewidth][c]{\includegraphics[width=0.88\linewidth,height=0.48\textheight,keepaspectratio]{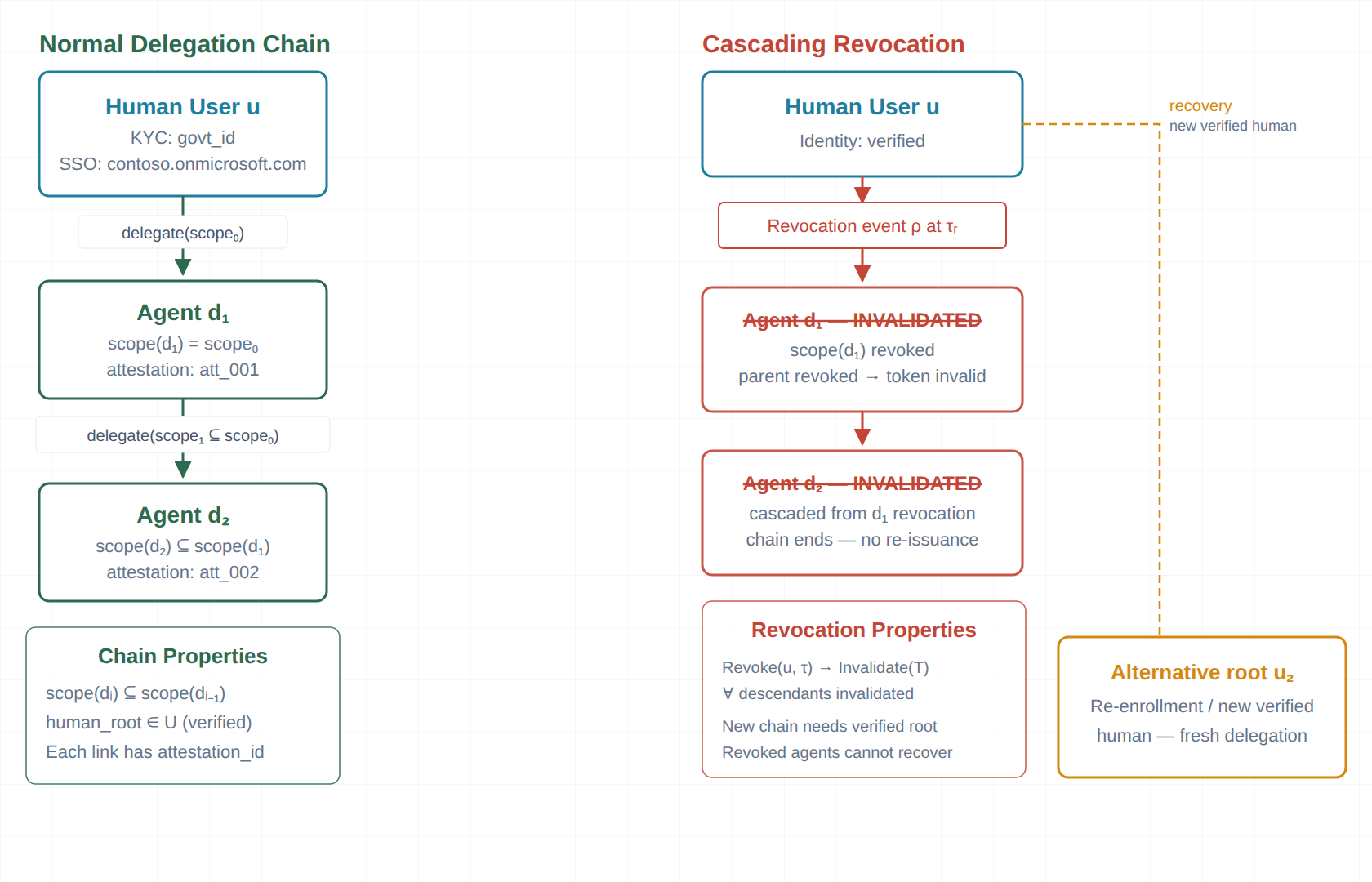}}
  \figcaption{Delegation Chains and Cascading Revocation}{Left: Normal delegation chain from human root $u$ through agents $d_1$, $d_2$ with monotonically non-expanding scope. Right: Revocation event $\rho$ at the human root invalidates descendants; the chain ends at revoked agents. Recovery requires a separately verified alternative root $u_2$ (re-enrollment or new human)---not re-issuance from invalidated agents.}
  \label{f:delegation}
\end{figure}

xChk access tokens (surface B) carry $\mathbf{D}$ via \path{xchk_delegation_chains} assembled from \path{attestation_completed} portfolio artifacts. The chain payload is \textbf{integrator-supplied} metadata on attestation create (\texttt{delegationChain} in the attestation request); the platform does not auto-walk the \texttt{parentAgentId} tree to mint token chains. By contrast, \texttt{parentAgentId} drives \textbf{cascading revocation}---a BFS tree walk on agent delete revokes descendant agents and their API keys (\S\ref{s:impl-status}).

Define a revocation event $\rho$ for entity $e$ at time $\tau_\rho$. The revocation propagates as:

\begin{equation}
  \text{Revoke}(e, \tau_\rho) \Rightarrow \{ \text{Invalidate}(\mathbf{T}) \;|\; \mathbf{T}.\mathit{entity\_id} \in \text{Descendants}(e) \}\label{e:revocation}
\end{equation}

where $\text{Descendants}(e)$ is the transitive closure of entities whose delegation chain contains $e$ as an ancestor. Production revokes descendant agents and their API keys when a parent agent is deleted.

\subsection{Multi-Entity Session Token}\label{s:session}

For workflows involving multiple agents operating under related delegation chains, a session token $\mathbf{T}_s$ is defined:

\begin{equation}
  \mathbf{T}_s = (\mathit{session\_id}, \{ \mathbf{D}_e \;|\; e \in \mathbf{E}_{\mathit{session}} \}, \mathit{scope}_s, S_s)\label{e:session}
\end{equation}

where $\mathit{scope}_s = \bigcap_e \mathit{scope}(\mathbf{D}_e)$ and $S_s = \min(\{S_e\})$. \texttt{POST /api/session-tokens} creates a session from up to 20 entities, computes intersection scope and composite trust score, and returns a \texttt{sessionId} embeddable in the \texttt{xchk\_session} claim via \texttt{POST /oauth/refresh-token}.

\FloatBarrier

\section{Implementation}\label{s:implementation}

A reference implementation of the xChk identity provider and attestation gateway is deployed at \url{https://in.xchk.io}. One documented relying party---\url{https://crabbyed.com} (Crabby Editor, Appendix~\ref{s:crabbyed})---exercises surface~A Login with xChk in production; broader third-party adoption is not yet demonstrated. The deployment covers OAuth 2.0 / OpenID Connect authorization code flow with PKCE, RS256-signed ID tokens and access tokens, a JWKS endpoint, \texttt{GET /oauth/userinfo}, and an OpenID Connect discovery document.

\subsection{Production Delivery Surfaces}\label{s:surfaces}

Throughout \S\ref{s:implementation}, \textbf{shipped} means the construct is available in production on at least one of the surfaces below. The tuples in \S\ref{s:model}--\S\ref{s:attestation} describe the intended token layout; the table in \S\ref{s:impl-status} states where each construct is emitted today.

\begin{table}[ht]
  \centering
  \footnotesize
  \begin{tabular}{@{}L{0.14\linewidth}L{0.28\linewidth}L{0.48\linewidth}@{}}
    \toprule
    \textbf{Surface} & \textbf{Who uses it} & \textbf{What ships} \\
    \midrule
    \textbf{A --- RP OIDC ID token} & Third-party apps via Login with xChk (\texttt{/oauth/authorize} $\rightarrow$ \texttt{/oauth/token}) & \texttt{xchk.portfolio}, \texttt{xchk.policy}, portfolio HMAC (\texttt{xchk.signature}), bilateral RP institutional claims, profile claims, \texttt{xchk\_entity\_type: human}, optional \texttt{xchk\_portfolio\_proofs} \\
    \textbf{B --- xChk access token} & First-party app via \texttt{/api/login}; refresh via \texttt{POST /oauth/refresh-token} & \texttt{xchk\_trust\_score}, \texttt{xchk\_portfolio\_summary}, \texttt{xchk\_attestations}, \texttt{xchk\_delegation\_chains}, optional \texttt{xchk\_session} \\
    \textbf{C --- Agent API} & Registered agents (API keys) & \texttt{POST /api/agents/authorize}, attestation create/respond, External Approval Policies, cascading revoke on agent delete, elevation grant verify \\
    \bottomrule
  \end{tabular}
  \caption{Production delivery surfaces.}
  \label{t:surfaces}
\end{table}

\textbf{Integrator surfaces.} Use \textbf{surface A alone} when a third-party RP only needs Login with xChk portfolio claims and bilateral consent (standard OIDC + nested \texttt{xchk}). \textbf{Add surface B} when the integrator must read post-approval \texttt{xchk\_attestations}, delegation chains, trust score, or session embeds without a full OIDC re-login (call refresh after approve). Use \textbf{surface C} for agent runtime governance (API keys, authorize/attest, elevation grants)---not as an OIDC bearer path for RPs.

\subsection{Implementation Status}\label{s:impl-status}

{\scriptsize
\begin{longtable}{@{}L{0.36\linewidth}ccL{0.22\linewidth}@{}}
  \caption{Implementation status of all constructs.}
  \label{t:status}\\
  \toprule
  \textbf{Construct} & \textbf{\S} & \textbf{Surf.} & \textbf{Status} \\
  \midrule
  \endfirsthead
  \multicolumn{4}{c}{\tablename\ \thetable{} (continued)} \\
  \toprule
  \textbf{Construct} & \textbf{\S} & \textbf{Surf.} & \textbf{Status} \\
  \midrule
  \endhead
  OAuth/OIDC + PKCE, RS256, JWKS, pairwise \texttt{sub} & 3.2 & A & Shipped \\
  \texttt{xchk.portfolio} + \texttt{consistency} / \texttt{behavioral} & 3.2 & A & Shipped \\
  \texttt{xchk.policy} + \texttt{xchk\_evidence\_policy} & 3.3 & A & Shipped \\
  Per-app artifact picker / disclosure & 3.2 & A & Shipped \\
  Fresh verification (WebAuthn / PIN) at consent & 3.3 & A & Shipped \\
  \texttt{xchk\_entity\_type} on OIDC ID token & 3.2 & A & Shipped --- \texttt{human} \\
  \texttt{issuer} + \texttt{issuerTier} provenance & 3.1 & A & Shipped \\
  Org trust ladder (DNS $\rightarrow$ KYB $\rightarrow$ officer binding) & 3.1 & A & Shipped \\
  \texttt{xchk\_portfolio\_proofs} & 3.2 & A & Shipped \\
  \texttt{xchk\_trust\_score}, \texttt{xchk\_portfolio\_summary} & 3.4 & B & Shipped \\
  \texttt{xchk\_attestations} embed & 4.2 & B & Shipped --- refresh after approve \\
  \texttt{xchk\_delegation\_chains} & 4.4 & B & Shipped \\
  Multi-entity \texttt{xchk\_session} & 4.5 & B & Shipped --- session-tokens + refresh \\
  \texttt{attestation\_completed} portfolio artifact & 4.2 & A & Shipped --- on next OIDC login \\
  \texttt{POST /api/agents/authorize} & 4.3 & C & Shipped \\
  Attestation hash chain + HMAC + elevation grant & 4.1--4.2 & C & Shipped \\
  Insufficient-approver validation & 4.2 & C & Shipped \\
  External Approval Policies (EAP) & 4.2 & C & Shipped \\
  Cascading agent revocation & 4.4 & C & Shipped \\
  Hermes guardrails E2E & 4.2 & C & Shipped --- integration; $\sim$50 patterns when Hermes runtime present (not vendored in xchk) \\
  Ravencoin OP\_RETURN anchor & 4.2 & C & Partial --- requires Ravencoin node \\
  Portfolio HMAC on ID token & 3.2 & A & Shipped --- consent snapshot \\
  Bilateral RP institutional portfolio evaluation & 3.3 & A & Shipped --- hard gate on release \\
  Production OAuth enforce (client tier + redirect URI) & 3.1 & A & Shipped --- DNS-verified redirect \\
  \texttt{GET /oauth/userinfo} & 3.2 & A & Shipped --- same nested \texttt{xchk} shape; live-evaluated (ID token is consent snapshot) \\
  \texttt{legally\_binding} attestation completion gate & 4.2 & C & Shipped --- requires \texttt{officer\_bound} \\
  Fleet pack batch import (\texttt{POST /api/agents/import}) & 4.3 & C & Shipped \\
  Agent plate id + registration venue (\texttt{plateState}) & 5.4 & C & Shipped --- cascade: explicit $\rightarrow$ owner KYC $\rightarrow$ \texttt{deployRegion} $\rightarrow$ FL \\
  Preregister invite domain gate & 3.1 & C/API & Shipped --- requires \texttt{entity\_verified} org \\
  Full non-human \texttt{entity\_type} on RP OIDC token & 3.2 & A & Future --- agents use surface C \\
  \bottomrule
\end{longtable}
}

\subsection{Identity Provider (OIDC)}\label{s:oidc}

\begin{itemize}
  \item Standard OAuth 2.0 authorization code flow with PKCE (S256)
  \item RS256-signed ID tokens with pairwise subjects; \texttt{GET /oauth/userinfo} returns the same nested \texttt{xchk} claims when \texttt{xchk\_portfolio} scope is granted, but \textbf{live-evaluates} portfolio and policy (ID token carries the consent-time snapshot)
  \item Extended claims: \path{xchk.portfolio}, \path{xchk.policy}, \path{xchk.signature}, \path{xchk.rp_institutional}, \path{xchk.subject_rp_policy}, \path{xchk_name_kind}, \path{xchk_legal_name_on_file}, \path{xchk_entity_type}
  \item Evidence policy: RPs transmit base64url JSON via \texttt{xchk\_evidence\_policy}; IdP evaluates against the user's portfolio and returns \texttt{xchk.policy} before consent---the RP still decides access
  \item Bilateral consent: subject-defined policy evaluated against the RP's institutional portfolio (domain control, KYB, officer binding); default requires DNS-verified \texttt{domain\_control}; portfolio release blocked when unsatisfied
  \item Portfolio integrity: HMAC-SHA256 over the canonical $\{\mathit{portfolio}, \mathit{policy}, \mathit{rpInstitutional}, \mathit{subjectRpPolicy}\}$ snapshot (\texttt{xchk-portfolio-v1} key id); verification procedure in Appendix~\ref{s:hmac}
  \item Production OAuth gates: OAuth client verification tier synced from org ladder; redirect URIs must match DNS-verified domains in production (localhost exempt in development)
  \item Fresh-verification enforcement: WebAuthn passkey or PIN re-verification during consent when the policy requires it
  \item Per-app disclosure: users choose which artifact types to share with each RP
\end{itemize}

\paragraph{Verification portfolio.} Assembled at token issuance from ten artifact types:

\begin{table}[H]
  \centering
  \footnotesize
  \begin{tabular}{@{}L{0.30\linewidth}L{0.26\linewidth}L{0.32\linewidth}@{}}
    \toprule
    \textbf{Artifact Type} & \textbf{Source} & \textbf{Assurance Signal} \\
    \midrule
    \nolinkurl{kyc_verified} & Didit (govt ID + liveness) & authLevel, provider, timestamp \\
    \nolinkurl{corporate_sso_linked} & Entra ID, Google Workspace & Tenant ID, domain, tenant age \\
    \nolinkurl{platform_identity_verified} & LinkedIn Identity Verification API & Verified name, ID category \\
    \nolinkurl{workplace_verified} & LinkedIn Workplace Verification & Organization name, date \\
    \nolinkurl{longitudinal_activity} & GitHub OAuth, LinkedIn & Account age, contributions, followers \\
    \nolinkurl{passkey_registered} & WebAuthn/FIDO2 authenticator & Credential ID, registration date \\
    \nolinkurl{helpdesk_pass} & xChk helpdesk sessions & Session ID, proof URL \\
    \nolinkurl{interview_pass} & xChk identity interview & Session ID, proof URL \\
    \nolinkurl{attestation_completed} & xChk attestation records & Attestation ID, optional \texttt{rvnTxId}, \texttt{sha256} \\
    \nolinkurl{behavioral_signals} & Device, network, location, time & Usual/new/unusual per dimension \\
    \bottomrule
  \end{tabular}
  \caption{Verification portfolio artifact types.}
  \label{t:artifacts}
\end{table}

\FloatBarrier

Every artifact names its \texttt{issuer} and \texttt{issuerTier} (\texttt{unverified}, \texttt{domain\_verified}, \texttt{entity\_verified}, \texttt{officer\_bound}). Organization enrollment adds DNS domain verification, KYB (Didit), and officer binding for institutional attestations.

\subsection{Attestation Gateway}\label{s:gateway}

\begin{itemize}
  \item Hermes guardrails E2E: xChk ships the integration (\texttt{examples/hermes-attestation-guardrails/}); dangerous-command classification delegates to Hermes \texttt{detect\_dangerous\_command} ($\sim$50 patterns when that runtime is installed---pattern catalog not vendored in xchk repos; demo fallback uses a small regex set)
  \item Hash-chained attestation records with HMAC signing
  \item Optional Ravencoin OP\_RETURN anchoring (v1/v2, 80-byte payload, \texttt{xCk} magic prefix)
  \item Progressive notification with approver pool expansion on timeout
  \item Notification channels: Slack, Telegram, SMS, WhatsApp
  \item Multi-approver escalation chains with policy-based routing; rejects create when approver pool is smaller than \texttt{requiredApprovals}
  \item Server-computed \texttt{bindingClass} (organizational / \texttt{legally\_binding}) from purpose templates and policy names; terminal \texttt{legally\_binding} completion requires org tier \texttt{officer\_bound}
  \item External Approval Policies (\texttt{POST /api/external-approval-policies}) for cross-domain attestation authorization
  \item Fleet pack batch import (\texttt{POST /api/agents/import}) with role/intent registry for multi-agent deployment
  \item WebAuthn passkey approver verification on attest-reply (enrollment PIN is not accepted for attestation unlock)
  \item \texttt{POST /api/agents/authorize} $\rightarrow$ \texttt{allow} $|$ \texttt{attest} $|$ \texttt{forbid}; signed elevation grant on approve
\end{itemize}

\paragraph{Jurisdiction of registration (production embodiment).}
Each registered agent carries a stable plate identifier (\texttt{TYP-BODY}, e.g.\ \texttt{OPS-YDDS}) and a U.S.\ registration venue (\texttt{plateState}: currently \texttt{FL}, \texttt{CA}, or \texttt{CO})---chosen the way contracts pick a forum, not a claim about where model weights sit. Venue resolution order: (1)~explicit operator-set \texttt{plateState}; (2)~human owner location from KYC document state; (3)~agent declared deploy / physical location (\texttt{deployRegion}, including common cloud-region hints); (4)~platform default \texttt{FL}. Provenance is stored as \texttt{plateStateSource} (\texttt{explicit} $|$ \texttt{owner\_kyc} $|$ \texttt{agent\_deploy} $|$ \texttt{default}). Deploying or acting in another jurisdiction (a ``foreign land'') does not extinguish home registration: place-of-operation rules may \emph{add} obligations; they do not erase the home venue recorded on the agent and its attestations. Attestation records retain plate and venue so an auditor can map a human approval to the regulatory surface under which the agent was licensed---without reading soft display names or API-key owner email. The production UI presents venue as state license-plate graphics (Florida, California, Colorado) so approvers see \emph{who} the agent is and \emph{where it is registered} in one glance; the cryptographic record stores the jurisdiction code, not the graphic.

Agents use API keys and the authorize/attest API---not OIDC tokens---for runtime governance. Human principals use Login with xChk for RP federation.

\subsection{API Authentication and Deployment}\label{s:api}

API middleware accepts Firebase ID tokens, xChk OAuth access tokens, Zendesk launch tokens, and SSO launch tokens. \texttt{/api/login} mints an \texttt{xchkAccessToken} for non-anonymous users; \texttt{POST /oauth/refresh-token} refreshes it with updated portfolio summary, attestation embeds, delegation chains, and optional session reference. \texttt{POST /api/session-tokens} creates multi-entity session records surfaced in the \texttt{xchk\_session} claim (\S\ref{s:session}).

\textbf{Deployment:} Node.js (Express), MongoDB, PM2 on Ubuntu 24.04, Nginx TLS termination; static frontend at \texttt{/var/www/xchk-app}.

\subsection{Micro-benchmark (public endpoints)}\label{s:microbench}

We measured end-to-end HTTPS latency from a client on the public Internet to production \texttt{in.xchk.io} (2026-07-11). Samples are wall-clock \texttt{curl} transfer times (TLS + server). Portfolio HMAC verify is a local microbench matching Appendix~\ref{s:hmac} (not network-bound).

\begin{table}[!ht]
  \centering
  \footnotesize
  \begin{tabular}{@{}lrrrL{0.38\linewidth}@{}}
    \toprule
    \textbf{Operation} & \textbf{n} & \textbf{p50} & \textbf{p95} & \textbf{Notes} \\
    \midrule
    \texttt{GET /.well-known/openid-configuration} & 50 & 159\,ms & 200\,ms & Public discovery \\
    \texttt{GET /oauth/jwks} & 50 & 161\,ms & 187\,ms & Public JWKS \\
    \texttt{POST /oauth/token} (invalid client) & 50 & 162\,ms & 191\,ms & Reject path; not a successful code exchange \\
    \texttt{POST /api/agents/authorize} (no API key) & 50 & 165\,ms & 213\,ms & Auth-fail path; not a scoped allow/attest decision \\
    Portfolio HMAC-SHA256 verify (local) & 5000 & 0.014\,ms & 0.031\,ms & Synthetic 10-artifact portfolio \\
    \bottomrule
  \end{tabular}
  \caption{Public-endpoint and local HMAC micro-benchmark (2026-07-11).}
  \label{t:microbench}
\end{table}

Successful authorize/attest and full OIDC code exchange latencies depend on credentials and human approval and are not reported here. Human-in-the-loop attestation remains dominated by approver response time (\S\ref{s:limitations}).

\section{Discussion}\label{s:discussion}

\subsection{Security Considerations}\label{s:security}

\textbf{Threat model (summary).} We consider: (i) an honest-but-curious RP that receives only consented claims and may attempt cross-RP correlation; (ii) a compromised modality issuer or stolen SSO session for a single enrollment channel; (iii) a stolen agent API key used to propose actions; (iv) IdP database compromise exposing stored portfolios and attestation records; (v) leakage of \texttt{ATTESTATION\_SIGNING\_SECRET} used for portfolio HMAC and attestation integrity. Pairwise \texttt{sub}, per-app disclosure, and multi-modality RP policies mitigate (i)--(ii). Surface C authorize/attest plus human approval and elevation grants mitigate (iii) for gated actions (stolen keys can still call APIs until revoked). (iv) and (v) are out of scope for RP-side JWT verification alone---self-hosted deployments must protect the signing secret; hosted deployments coordinate secret distribution out of band. RS256 JWT validation without portfolio HMAC does not detect portfolio substitution inside an otherwise valid ID token.

\textbf{Token forgery:} The IdP signs tokens with RS256 JWT signatures. RPs verify against the published JWKS endpoint. Portfolio packages additionally carry an HMAC (\texttt{xchk.signature}) over the disclosed snapshot (Appendix~\ref{s:hmac}; verifier in Appendix~\ref{s:hmac-verifier}). Self-hosted deployments share the attestation signing secret with integrators who verify portfolio integrity independently of ongoing IdP read access; hosted deployments at in.xchk.io coordinate secret distribution out of band.

\textbf{Replay attacks:} OIDC \texttt{nonce} and \texttt{exp} prevent token replay. The attestation gateway binds \texttt{authorizeSessionId} to action hashes.

\textbf{Modality compromise:} If a single modality is compromised (e.g., a leaked corporate SSO session), remaining modalities continue to provide assurance. RPs may require multiple artifact types for sensitive operations.

\textbf{Session integrity:} The attestation chain enforces temporal ordering via \texttt{chain\_hash}. Ordering anomalies may indicate session takeover or replay.

\textbf{Blockchain anchoring:} When configured, Ravencoin provides a timestamped commitment to the attestation hash. The blockchain is not an execution layer; records remain in the IdP database and are verifiable against the anchor.

\subsection{Privacy Considerations}\label{s:privacy}

BYOI limits exposure: users choose enrollment and per-RP disclosure; tokens carry assurance indicators, not raw KYC documents. Pairwise \texttt{sub} prevents cross-RP correlation. The per-app artifact picker lets users withhold artifact types---\texttt{xchk.policy} is evaluated on the disclosed subset only. \texttt{xchk\_portfolio\_proofs} issues expiring, audience-bound proof URLs. Users may invoke \textbf{Forget Me} erasure, which invalidates proof links early. Name provenance (\texttt{xchk\_name\_kind}) lets RPs accept pseudonymous-but-accountable identities (\texttt{registered\_alias} with KYC on file) without receiving the legal name. Session photos and likeness data are acknowledged PII-lite exceptions stored for verification, not exported as raw biometrics in tokens.

\subsection{Relationship to Behavioral Governance}\label{s:governance}

As in \S\ref{s:agentbound}, AgentBound~\citep{agentbound} governs behavioral correctness while xChk governs identity assurance and human-in-the-loop attestation; AgentBound review obligations can flow through xChk's attestation gateway, and governance receipts plus attestation records together bind behavioral and identity-layer decisions.

\subsection{Limitations}\label{s:limitations}

\begin{itemize}
  \item \textbf{RP adoption:} RPs must parse custom claims and evaluate policies; standard OIDC libraries do not do this automatically.
  \item \textbf{Delivery surfaces:} Extended claims (\texttt{xchk\_attestations}, delegation chains, session) require the xChk access-token refresh path (\S\ref{s:surfaces}), not standard RP OIDC ID tokens alone.
  \item \textbf{External validation:} One documented relying party (\url{https://crabbyed.com}, Appendix~\ref{s:crabbyed}) exercises surface~A in production; broader integrator adoption remains future work.
  \item \textbf{Enrollment UX:} Modalities differ in self-service depth---passkey registration, corporate SSO linking, and helpdesk/interview flows are user-initiated; org KYB, officer binding, and some professional-network paths require operator configuration.
  \item \textbf{Commitment-class classifier:} Attestation \texttt{bindingClass} uses server heuristics and policy templates; optional org-configured rules or LLM classifiers are not yet deployed platform-wide.
  \item \textbf{Trust score advisory role:} \texttt{xchk\_trust\_score} is advisory; RPs define thresholds over portfolio signals.
  \item \textbf{Blockchain anchoring:} Requires a connected Ravencoin node with funded wallet when enabled.
  \item \textbf{Latency:} Human-in-the-loop attestation introduces variable delay---unsuitable for hard real-time loops.
  \item \textbf{Evaluation:} No formal user study; \S\ref{s:microbench} reports public-endpoint and local HMAC micro-benchmarks only. Successful authorize/attest timings remain future work.
\end{itemize}

\section{Conclusion}\label{s:conclusion}

We presented xChk, a BYOI architecture for OAuth/OIDC that encodes heterogeneous identity assurance as machine-readable portfolio claims and leaves access sufficiency to each relying party. We described an architectural model with a three-way separation---user controls disclosure, IdP transports claims (and optionally evaluates an RP-supplied policy for consent), RP adjudicates access---extended with bilateral evaluation of the RP's institutional portfolio before subject portfolio release. A reference implementation at in.xchk.io ships on three surfaces (\S\ref{s:surfaces}); surface~A is documented with one production RP (Appendix~\ref{s:crabbyed}). Agent governance, optional Ravencoin anchoring, and full non-human OIDC entity types follow \S\ref{s:impl-status} Partial/Future rows.

\section{Data and Code Availability}\label{s:availability}

The reference implementation is deployed at \url{https://in.xchk.io}. Source code is maintained in private repositories and is \textbf{not publicly available} at the time of writing. Third parties can interact with the live system without a local checkout: OpenID Connect discovery (\texttt{/.well-known/openid-configuration}), the OAuth playground (\url{https://in.xchk.io/oauth/playground.html}), the Login with xChk integrator guide (\url{https://in.xchk.io/login-with-xchk.html}), and the public portfolio HMAC verifier (\url{https://in.xchk.io/verify-portfolio-hmac.cjs}, Appendix~\ref{s:hmac-verifier}). The paper PDF with embedded architecture diagrams is at \url{https://in.xchk.io/xchk-paper-with-diagrams.pdf}. No proprietary dataset is required to exercise the OAuth portfolio flow against production; integrators may register OAuth clients on the hosted service.

\appendix
\section{Portfolio HMAC Verification}\label{s:hmac}

RPs verifying surface A ID tokens should recompute the portfolio HMAC after RS256 JWT validation:
\begin{enumerate}
  \item Extract portfolio fields from the nested \texttt{xchk} claim: \path{payload.xchk.portfolio}, \path{payload.xchk.policy}, \path{payload.xchk.rp_institutional}, and \path{payload.xchk.subject_rp_policy} (paper notation \texttt{xchk.portfolio}, etc.). Use top-level \path{payload.xchk_signature} or \path{payload.xchk.signature} for the HMAC digest.
  \item Build canonical UTF-8 JSON with recursively sorted object keys:
\end{enumerate}

\begin{lstlisting}
{
  "policy": { "satisfied": true, "missing": [] },
  "portfolio": { "artifacts": [], "consistency": {}, "behavioral": {} },
  "rpInstitutional": { "artifacts": [] },
  "subjectRpPolicy": { "satisfied": true, "missing": [] }
}
\end{lstlisting}

\begin{enumerate}
  \setcounter{enumi}{2}
  \item Compute HMAC-SHA256 using the first 32 hex characters of \texttt{SHA256(ATTESTATION\_SIGNING\_SECRET)} as the key.
  \item Compare the hex digest to \texttt{xchk.signature} / \texttt{xchk\_signature}. Mismatch indicates tampering or token/portfolio substitution.
\end{enumerate}

Key id \texttt{xchk\_signing\_key\_id: xchk-portfolio-v1} identifies the signing scheme. Full integrator steps: \url{https://in.xchk.io/login-with-xchk.html}. A runnable verifier matching this procedure is Appendix~\ref{s:hmac-verifier}.

\section{Documented RP Integration (crabbyed.com)}\label{s:crabbyed}

\textbf{Crabby Editor} (\url{https://crabbyed.com}) is a verification-first web application that uses Login with xChk for continuity after identity verification---not as a generic social-login replacement. It is the only RP integration documented in this paper at production scale; it shares operational ownership with the xChk deployment and should be read as a \textbf{reference integration}, not independent ecosystem adoption.

\paragraph{OAuth client.} Registered OAuth client \texttt{client\_id=crabbyed.com} with production redirect URI on the Crabby Editor origin. Login entry point: \url{https://crabbyed.com/login.html}.

\paragraph{Authorization flow (surface~A).}
\begin{enumerate}
  \item Browser initiates OAuth 2.0 authorization code flow with PKCE (S256) against the xChk authorize endpoint (\texttt{/oauth/authorize} on \url{https://in.xchk.io}).
  \item Scopes: \texttt{openid}, \texttt{profile}, \texttt{email}, \texttt{xchk\_portfolio}; optional \texttt{xchk\_portfolio\_proofs}.
  \item User authenticates on xChk, selects per-app artifact disclosure, and completes bilateral consent (subject policy vs RP institutional portfolio).
  \item Crabby Editor backend exchanges the code at \texttt{POST /oauth/token}, verifies the RS256 ID token (discovery + JWKS), and optionally verifies the portfolio HMAC (Appendix~\ref{s:hmac}).
  \item Backend mints a Crabby session from \texttt{sub} / \texttt{email} and stores disclosed portfolio claims for policy evaluation.
\end{enumerate}

\paragraph{RP policy evaluation.} Crabby Editor reads nested \texttt{xchk.portfolio} artifacts and \texttt{xchk.policy} from the consent-time ID token snapshot. Access decisions are RP-local: the IdP displays policy satisfaction on the consent screen but does not grant or deny Crabby access. Typical gates inspect artifact types (e.g.\ helpdesk pass, interview pass, KYC linkage) via the RP's own sufficiency function---the BYOI pattern from \S\ref{s:byoi}.

\paragraph{Integration surface area.} Beyond standard OIDC (discovery, JWKS, PKCE, token exchange), the RP adds: parsing nested \texttt{xchk} claims, optional portfolio HMAC verification, and application policy over artifact types. Integrators can reproduce the flow using the hosted OAuth playground (\url{https://in.xchk.io/oauth/playground.html}) before deploying a callback. Step-by-step guide: \url{https://in.xchk.io/login-with-xchk.html}.

\paragraph{What this appendix does not claim.} No controlled A/B measurements are reported beyond the public-endpoint micro-benchmark in \S\ref{s:microbench}; no second independent RP is documented; agent surfaces B/C are exercised on in.xchk.io but not through Crabby Editor's OIDC login path.

\section{Portfolio HMAC Verifier (Node.js)}\label{s:hmac-verifier}

Runnable verifier matching Appendix~\ref{s:hmac} and production \texttt{portfolioSignatureUtils.js}. Also published at \url{https://in.xchk.io/verify-portfolio-hmac.cjs}.

\begin{lstlisting}[basicstyle=\ttfamily\scriptsize]
// verify-portfolio-hmac.cjs --- xchk-portfolio-v1
const crypto = require('crypto');

function portfolioSigningKey(secret) {
  return crypto.createHash('sha256').update(secret)
    .digest('hex').slice(0, 32);
}

function sortKeysDeep(value) {
  if (Array.isArray(value)) return value.map(sortKeysDeep);
  if (value && typeof value === 'object') {
    return Object.keys(value).sort().reduce((acc, key) => {
      acc[key] = sortKeysDeep(value[key]);
      return acc;
    }, {});
  }
  return value;
}

function verifyPortfolioSignature({
  portfolio, policy, rpInstitutional, subjectRpPolicy,
  signature, secret,
}) {
  const key = portfolioSigningKey(secret);
  const canonical = JSON.stringify(sortKeysDeep({
    portfolio,
    policy: policy ?? null,
    rpInstitutional: rpInstitutional ?? null,
    subjectRpPolicy: subjectRpPolicy ?? null,
  }));
  const expected = crypto.createHmac('sha256', key)
    .update(canonical).digest('hex');
  if (!signature || expected.length !== signature.length) return false;
  return crypto.timingSafeEqual(
    Buffer.from(signature, 'hex'),
    Buffer.from(expected, 'hex'));
}

module.exports = {
  verifyPortfolioSignature, portfolioSigningKey, sortKeysDeep
};
\end{lstlisting}

Full file (with usage comments): \url{https://in.xchk.io/verify-portfolio-hmac.cjs}.

\bibliography{paper}

\end{document}